\documentclass{vgtc}                          

\ifpdf
  \pdfoutput=1\relax                   
  \usepackage{graphicx}                
  \DeclareGraphicsExtensions{.pdf,.png,.jpg,.jpeg} 
\else
  \usepackage{graphicx}                
  \DeclareGraphicsExtensions{.eps}     
\fi%

\graphicspath{{figures/}{pictures/}{images/}{./}} 

\usepackage{microtype}                 
\PassOptionsToPackage{warn}{textcomp}  
\usepackage{textcomp}                  
\usepackage{mathptmx}                  
\usepackage{times}                     
\usepackage{cite}                      
\usepackage{tabu}                      
\usepackage{booktabs}                  

\usepackage{paralist}
\usepackage{scalerel}

\usepackage[normalem]{ulem}
\useunder{\uline}{\ul}{}

\usepackage[dvipsnames]{xcolor}

\onlineid{1122}

\vgtccategory{Theory or Model}

\vgtcinsertpkg

\title{Boundary Blending: Reconsidering the Design of Multi-View Visualizations}

\newcommand*\samethanks[1][\value{footnote}]{\footnotemark[#1]}

\author{
\parbox{0.96\textwidth}{Maoyuan Sun\thanks{e-mail: \{smaoyuan, ashaikh2, yma2, dakoop, alhoori\}@niu.edu.} \hspace{14mm} Abdul Rahman Shaikh\samethanks  \hspace{14mm} Yue Ma\samethanks \hspace{14mm} David Koop\samethanks \hspace{14mm} Hamed Alhoori\samethanks}\\
\parbox{0.92\textwidth}{\hspace{0.35\textwidth}\scriptsize{$^{*}$Northern Illinois University}} \\
\vspace{-10mm}
} 

\abstract{
Multiple-view visualizations (MVs) have been widely used for visual analysis.
Each view shows some part of the data in a usable way, and together multiple views enable a holistic understanding of the data under investigation.
For example, an analyst may check a social network graph, a map of sensitive locations, a table of transaction records, and a collection of reports to identify suspicious activities.
While each view is designed to preserve its own visual context with visible borders or perceivable spatial distance from others, the key to solving real-world analysis problems often requires ``breaking" such boundaries, and further integrating and synthesizing the data scattered across multiple views.
This calls for \textit{blending} the boundaries in MVs, instead of simply breaking them, which brings key questions: what are possible boundaries in MVs, and what are design options that can support the boundary blending in MVs?
To answer these questions, we present three boundaries in MVs: 1) \textit{data} boundary, 2) \textit{representation} boundary, and 3) \textit{semantic} boundary, corresponding to three major aspects regarding the usage of MVs: encoded information, visual representation, and interpretation.
Then, we discuss four design strategies (\textit{highlighting}, \textit{linking}, \textit{embedding}, and \textit{extending}) and their pros and cons for supporting boundary blending in MVs.
We conclude our discussion with future research opportunities.
} 


\begin{document}


\firstsection{Introduction}

\maketitle


Multiple-view visualizations (MVs) is a commonly used visual analysis technique in many fields (e.g., bioinformatics \cite{steinberger2011context}, cybersecurity \cite{zhang2015visualizing}, finance \cite{bitextract, ko2012marketanalyzer}, healthcare \cite{sultanum2022chartwalk}, and text analysis \cite{dou2012leadline, jigsaw}).
MVs offer diverse perspectives on datasets, which allows users to find hidden patterns, relationships, and trends effectively \cite{roberts2007state}.
When using MVs, visualized data is not always spatially near each other, so users may need to explore related data from multiple views.
While each view offers a specific, coherent visual context, users may require related data pieces, instead of separated ones from different contexts (i.e., multiple views), to make sense of the data.

When related data pieces are connected, they can potentially form useful chains of information that act as a key to solving analytical problems. For example, suppose Amy, an analyst, is investigating possible threats from a collection of intelligence reports.
She uses natural language processing techniques to extract useful named entities (e.g., people names, locations, and organizations) from the text and plots them in multiple views (e.g., a list, a social network graph, and a map).
By checking each individual visualization, Amy can find some interesting data pieces (e.g., locations with a high incidence of reported activities, a few isolated network communities, and sensitive named organizations).
However, to develop her findings, Amy must relate, connect, and synthesize these data pieces, potentially breaking and blurring the boundary of these visualizations.

This shows one typical example of how MVs are commonly used.
It reveals that, for a variety of real-world analytical problems, the visible boundaries of views (i.e., borders and spatial distance, which are often designed and applied for the purpose of visual separation) are not that important, but, instead, often become barriers for users to perform visual analysis.
With the presence of visible boundaries, users have to overcome gaps between possibly related data pieces in MVs, by ``breaking" the boundaries. Such gaps not only appear spatially (i.e., layout) corresponding to the perception of MVs, but also logically regarding the interpretation of MVs. 
To address these gaps, it may be necessary to bridge related pieces of data that are contained within multiple boundaries. 
It goes beyond breaking the boundaries and involves an attempt to blend them together to create an integrated sensemaking space.

This critical challenge of designing MVs for supporting visual analysis prompts several essential questions: What are the possible boundaries in MVs? 
Can we break these boundaries and blend them further to enhance visual sensemaking? 
And, how does boundary blending influence the design of MVs? 
To answer these questions, we identify and present three possible boundaries in MVs: 1) \textit{data} boundary, 2) \textit{representation} boundary, and 3) \textit{semantic} boundary, corresponding to three primary aspects regarding the usage of MVs: encoded information, visual representation, and interpretation, respectively.
Based on them, we discuss four major design strategies (\textit{highlighting}, \textit{linking}, \textit{embedding}, and \textit{extending}) and their pros and cons for boundary blending in MVs.
Our insights on boundary blending can enrich the design space of MVs, offering valuable insights to inform more effective designs that support visual analysis.








\section{Boundaries in Multi-View Visualizations}
\label{boundary-def}


As mentioned above, there are three possible types of boundaries in MVs.
The first is the data boundary, which pertains to the information used to build MVs. The second is the representation boundary, which emphasizes visual separation, such as separating one view from another. The third is the semantic boundary, which reflects the interpretation of data in MVs.

\subsection{The Data Boundary}
The data boundary pertains to the information used for building MVs.
Since the design of visualizations heavily relies on user tasks \cite{munzner2014visualization},  each view within the context of data analysis encodes some part(s) of data to support certain analysis tasks \cite{sun2021towards}. 
Such encoded data provides the foundation for building MVs, which reveals the data in some useful way. 
Thus, possible overlaps in data must be considered, as different views within MVs can encode non-overlapped data, fully overlapped data, or partially overlapped data. This results in varying levels of clarity for the data boundary within MVs.

The encoded different data parts can serve as an underlying boundary for distinguishing between views.
For example, for the same set of textual documents, the extracted people’s names can be shown in a node-link diagram as a social network, and the extracted locations are displayed on a map.
These two different parts of data, encoded within the two views, establish a clear boundary that separates different types of information from the dataset.

In the case of encoding the same part(s) of data in different views, consider the example of a car dataset \cite{car-dataset}. One view can present all the data in a table to enable users to view detailed information about each car. Alternatively, the data can be visualized in a dimension reduction plot, using multidimensional scaling to reveal the similarities among the cars based on all attributes.
Both representations are produced by considering all the data within the dataset. 
Thus, there is no distinct data boundary between the two views, despite their visual differences.

In the case of partially overlapped data, consider the previous car dataset \cite{car-dataset} as an example. Two different bar charts can be created by using different attributes of the cars. For example, one bar chart may display the horsepower of all cars, while the other may show the weight.
They use the names of vehicles for bars (i.e., each car is represented as a separate bar). This overlap blurs the data boundary between the two views.

The clarity of the data boundary within MVs varies depending on the level of overlap in the data, ranging from a  \textit{clear} boundary to a \textit{blurred} one and even to \textit{no} boundary at all. This characteristic fundamentally supports boundary blending within MVs, which can potentially be achieved by manipulating the encoded data.

\subsection{The Representation Boundary}

\begin{figure}[tb]
  \centering
  \includegraphics[width=\columnwidth,keepaspectratio]{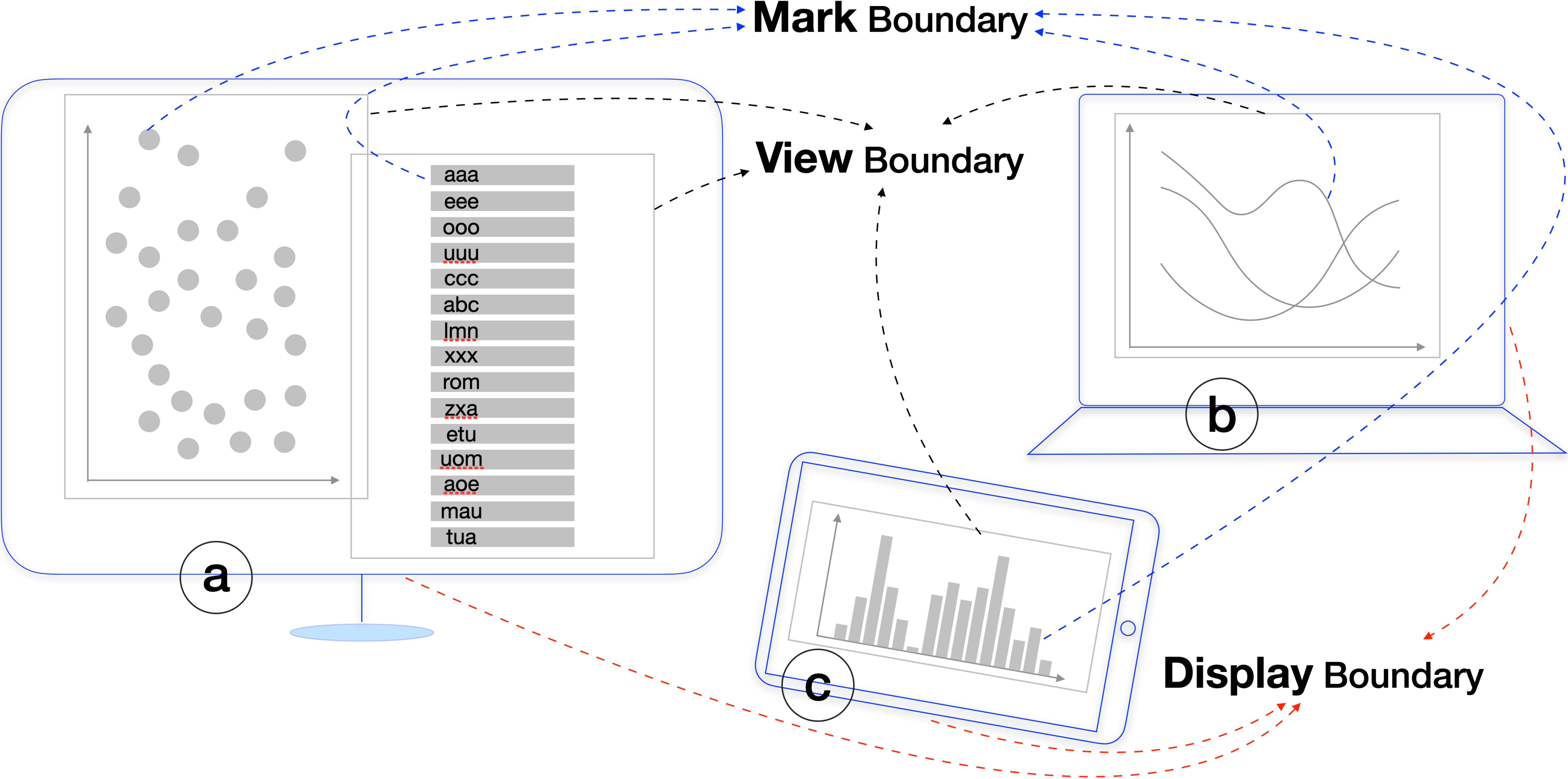}
\vspace{-6.5mm}
  \caption{Examples of three levels of representation boundary when using MVs in a multi-display environment. 
  (a) presents a large display showing a scatterplot and a list. 
  (b) is a laptop that shows a line chart.
  (c) reveals a tablet on which a bar chart is presented.}
  \label{fig-representation-bound}
\vspace{-5.5mm}
\end{figure}

The representation boundary pertains to the visual separation among different components involved in the usage of MVs.
It, in a broad sense, covers three levels of components, which corresponds to a hierarchical organization, including \textit{visual marks}, \textit{views}, and \textit{displays}.
Specifically, collections of spatially organized marks form views, sitting on one or multiple displays.
This spatial arrangement forms a hierarchical way of organization, where the higher level serves as the container for the lower level.
Figure \ref{fig-representation-bound} illustrates examples of the representation boundary for the three levels of components.

The boundary of visual marks separates one visual element from another.
This is often revealed as distinguishable visual marks (i.e., non-overlapped marks, or marks with different shapes or colors).
As each visualization can be decomposed as a collection of semantically coherent marks \cite{wilkinson2012grammar, munzner2014visualization}, visual separation of the marks is critical for users to understand visualizations. 
Moreover, if marks are not visually distinguishable, visual clutters can appear, which, for visualization design, is often attempted to be avoided.

The boundary of a view helps with preserving combined sets of marks as a coherent collection so that users can discriminate one view from another.
It attempts to support that each view can have its own distinct visual context, which may include unique graphical elements (i.e., visual marks), layout, or interaction methods.
Such a boundary is commonly revealed as borders outside a view, or empty space in-between views.
The former brings an enclosed space, which sometimes can be resized via user interactions. 
Different enclosed spaces help users recognize different views, even when the views overlap or are overlaid as composite visualizations \cite{javed2012exploring}.
The latter uses spatial distance to separate different views, and such a distance can sometimes be adjusted by users (e.g., moving a view).

The boundary of displays is the physical borders (i.e., bezels) of display devices (i.e., laptops, tablets, monitors, and wall displays).
Compared to a single display, a multiple-display environment offers more pixels to render and 
accommodate a greater number and variety of visualizations, 
making it ideal for using MVs for data analysis \cite{chung2014visporter, chung2015four, chung2018savil}.
With these physical borders, users can easily 
distinguish between displays, which may be used for supporting different analysis tasks, such as investigating a big social network graph on a large, high-resolution display, examining locations from a map on a desktop monitor, and reading textual documents on a laptop or tablet.

The three representation boundaries play a critical role in visual separation, which enables users to recognize different visual marks, views, and displays.
However, they also become barriers for users to develop a holistic understanding of the data in MVs, as the related data can be visually separated by the representation boundary.
Consequently, there is a need to build a global context that 
encompasses all the views across multiple displays, which potentially requires breaking and blending the visible representation boundaries.

\subsection{The Semantic Boundary}

\begin{figure}[tb]
  \centering
  \includegraphics[width=\columnwidth,keepaspectratio]{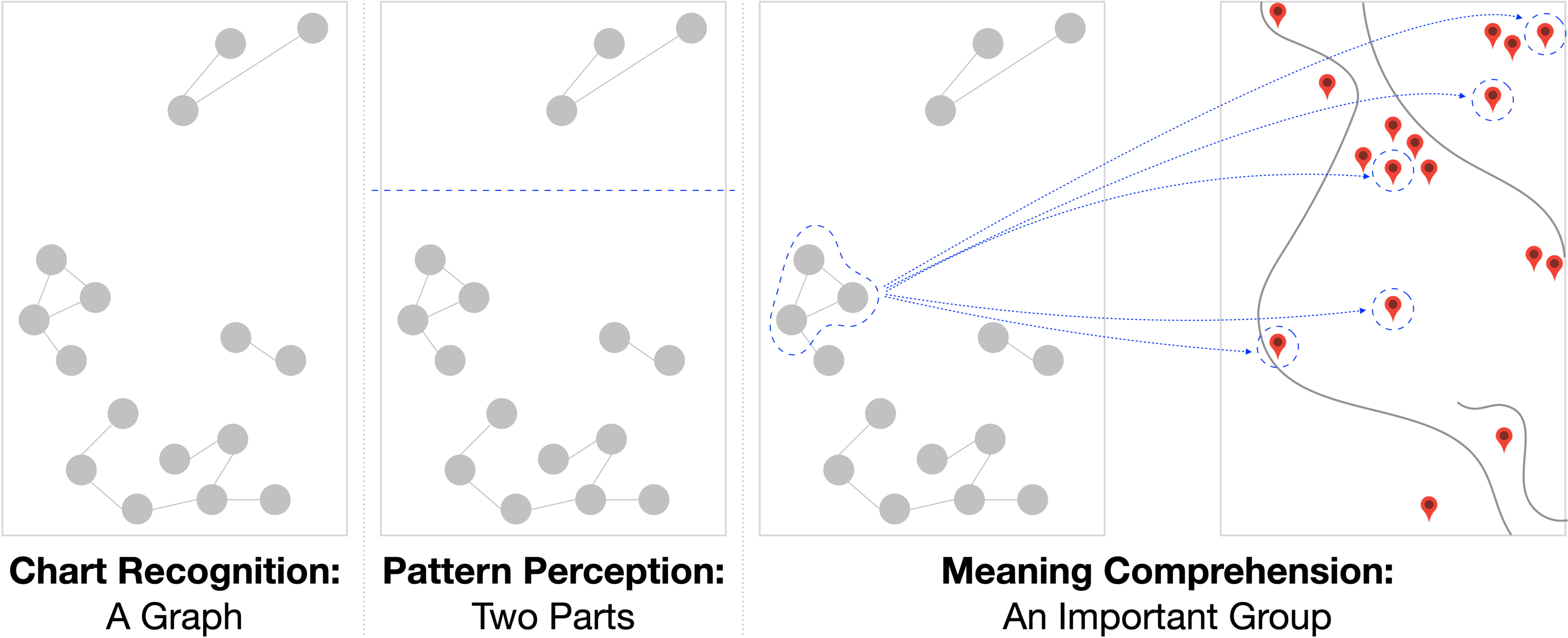}
\vspace{-6.5mm}
  \caption{Examples of multiple aspects of semantic boundary when interpreting views.}
  \label{fig-semantic-bound}
\vspace{-5.5mm}
\end{figure} 

The semantic boundary regards the differences in the interpretation and understanding of MVs.
This concept has been studied and used in the field of computer vision for image segmentation \cite{li2021multitask, acuna2019devil, marin2019efficient}.
In the context of data analysis with MVs, it is associated with the cognitive processes involved in analyzing and making sense of the data presented in MVs.
This boundary highlights unique insights, meanings, and knowledge obtained from MVs, and they may vary based on the user's prior knowledge, mental models, and analysis context.
Thus, based on possible interpretation of MVs, the semantic boundary can involve three aspects: 1) \textit{chart recognition}, 2) \textit{visual pattern perception}, and 3) \textit{meaning comprehension}.
Figure \ref{fig-semantic-bound} shows examples of them.

Regarding chart recognition, the semantic boundary refers to that users can clearly differentiate views based on the type of visualizations (i.e., recognizing a bar chart, map, matrix, list, or graph).
This heavily relies on the human's visual perception system \cite{cornsweet2012visual}, rather than the underlying data.
In addition, this boundary can be easily set, without taking much cognitive effort from users.
For example, as is shown in Figure \ref{fig-semantic-bound}, one can easily recognize that the leftmost visualization is a graph, while the rightmost one is a map.

For visual pattern perception, the semantic boundary serves as a means of discriminating patterns (i.e., clusters, trends, and outliers).
To perceive different patterns, the semantic boundary can be established based on visual contrast (e.g., different colors) and spatial arrangement (e.g., the layout or relative positions of visual elements).
Unlike chart recognition, which considers all visual elements in a view as a whole to get the type of chart, this type of semantic boundary delves deeper into views.
It goes beyond simply recognizing different charts to more detailed, visible patterns inside views.
However, similarly, the meanings of the underlying data of the perceived visual patterns do not receive much attention, so it is also driven by the human's visual perception system.

For meaning comprehension, the semantic boundary highlights considering the meanings of visualized data in MVs.
It not only involves the human visual perception system, but also actively engages the human reasoning system \cite{johnson2010mental}.
Unlike the other two aspects, this type of semantic boundary is set by the interpreted meanings of visualized data, which may or may not match perceived visual patterns.
This is the key to solving analysis problems.
For example, in Figure \ref{fig-semantic-bound} (right), a group of nodes in the graph is considered important and different from the rest, as they are all related to a few locations on the map, where sensitive activities were reported. 
The spatial arrangement does not necessarily separate these nodes from others, but the interpreted meanings clearly differentiate them from the rest.

Among the three aspects of the semantic boundary, chart recognition 
is the least concerned with the encoded data in MVs, whereas meaning comprehension places the greatest emphasis on the encoded data.
Moreover, they require different amounts of cognitive effort from users and involve different insights from visualizations \cite{north2006toward}.
The meaning comprehended from MVs is the key to sensemaking of data, which calls for breaking the representation boundary and crossing the data boundary.



\subsection{Relationships among the Three Boundaries}
The three boundaries (i.e., data, representation, and semantic), discussed before, reveal different focuses regarding the usage of MVs.
While they are not always independent, they can interact with each other.
Based on analysis tasks, certain data is picked, which is further visualized and organized on displays.
Such selected data determines the data boundary and their correspondingly generated views.
The spatial arrangement of these views drives the representation boundary.
Moreover, the semantic boundary results from a meaningful synthesis that considers encoded data and visible representations.
In turn, it further impacts the data and representation boundary.
For example, after identifying a suspicious group of people from a graph who all frequently traveled to three sensitive locations, a user moves them to the center of her current focus of view, and further investigates more details about them (e.g., bank accounts, transactions, and affiliated institutions) in several other views.
By doing so, both data and representation boundary get changed, as more data and views are pulled in, and the existing layout (i.e., the position of nodes in a graph) is modified.
Thus, during a sensemaking process, the three boundaries impact each other and update dynamically.



\section{Blending Boundaries in Multi-View Visualizations}
\label{sec-blending}

Separation is important for performing visual analysis.
However, to support sensemaking of data using MVs, logically connecting and synthesizing different pieces of related data, deriving meanings from such data, and developing usable insights are more crucial.
Thus, it calls for shifting the focus of using these boundaries in MVs from a simple separation toward a meaningful integration and connection.
This potentially requires, if necessary, breaking existing boundaries, and blending them to form an integrated and connected sensemaking space for solving complex data analysis problems.

\subsection{Defining Boundary Blending}
The concept of blending has been studied in cognitive science, which has been developed and explained as the blending theory \cite{coulson2005blending, fauconnier2003conceptual, coulson2001blending, grady1999blending}.
This theory focuses on conceptual blending.
It describes meaning construction as multiple cognitive models being integrated to form a blended mental space based on underlying structures and relations of given inputs.
For example, in language processing, given multiple documents, readers can develop some cognitive model for each based on their understanding of the text, then connect them according to comprehended relations (e.g., sharing the same person names and dates, or reporting related activities), and further develop a blended cognitive model, which corresponds to a comprehensive understanding of these documents (i.e., a summary).

While such a meaning-oriented blending is one key concern for sensemaking of data, especially considering the semantic boundary, boundary blending in MVs covers more than this.
As our identified boundaries (see Section \ref{boundary-def}) in MVs includes three aspects: encoded data, visible representation, and interpretation, we consider the major characteristics of boundary blending in MVs as:

\vspace{1.2mm}
\begin{compactitem}
  \item \textit{Integrated data space}: the data used for building MVs should be treated and constructed as an integrated data pool, instead of left as separated pieces, so that useful data can be efficiently retrieved and dynamically derived, on demand.
  \vspace{1mm}
  
  \item \textit{Connected representation space}: the representations of MVs should be reasonably linked, regardless of their bounded visual context (i.e., in a view or a display), so that users can view and explore chains of visualized data in a coherent context.
  \vspace{1mm}
  
  \item \textit{Synthesized knowledge space}: the interpretation of MVs should be aimed to be maintained as a synthesized web of knowledge, so that users can effectively navigate through their developed interpretation as usable knowledge to solve analysis problems.
\end{compactitem}
\vspace{1.2mm}

All together, the boundary blending in MVs highlights a process, which aims to form a synthesized knowledge space, for sensemaking of data, via the scaffold of using a connected representation space to investigate an integrated data space.
Therefore, for the boundary blending in MVs, toward forming a synthesized knowledge space is its key \textit{intention}, which is grounded in an integrated data space (considered as its \textit{foundation}), with the critical means of a connected representation space (considered as its \textit{implementation}).

\subsection{Designing Boundary Blending}

Our discussion on designing the boundary blending in MVs focuses on the representation boundary.
The data boundary relies on analysis problems, as different problems often need different data to develop corresponding solutions.
The data boundary does not directly interact with users in a visual analysis process.
Instead, it is wrapped by certain visual representations, so it indirectly impacts the perception of MVs.
Moreover, to enable an integrated data space, data fusion techniques can be used, which have been studied \cite{meng2020survey, castanedo2013review, bleiholder2009data}.  
For the semantic boundary, it emphasizes the interpretation of MVs, which cannot be directly manipulated by visualization designers.
However, to achieve the intention of the boundary blending, discussed above, the representations of MVs significantly impact users' interpretation.
Thus, considering its importance while using MVs, we focus on the design of blending the representation boundary.

\begin{figure}[tb]
  \centering
  \includegraphics[width=\columnwidth,keepaspectratio]{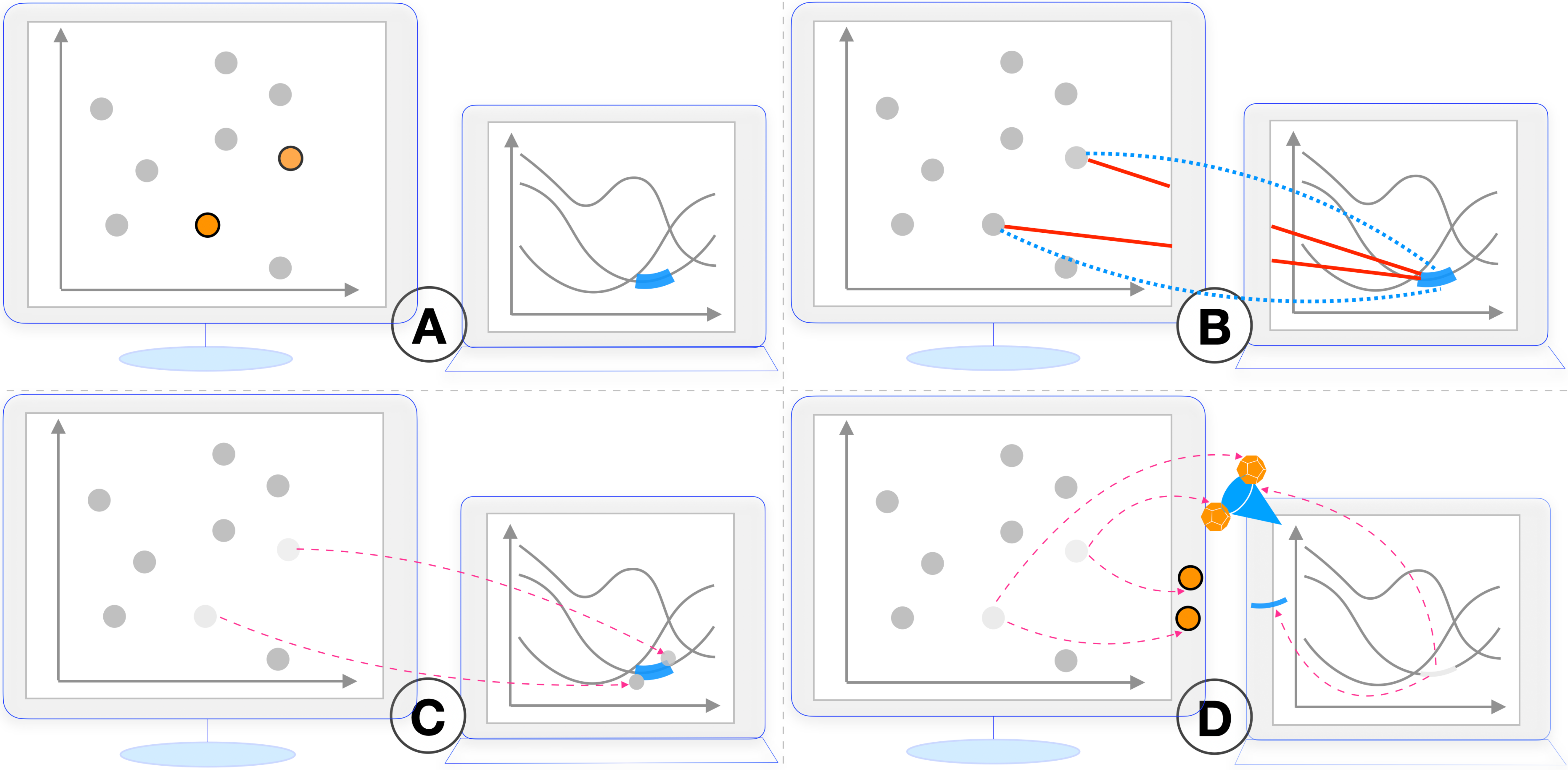}
  \vspace{-6.5mm}
  \caption{Examples of blending the representation boundary of two views on two displays (i.e., a desktop monitor and a laptop) with four different design strategies: (A) highlighting, (B) linking, (C) embedding, and (D) extending.
  Note: in (B), red lines show on-display visual links, while blue dashed lines are AR-rendered lines in the physical space.
  In (D), orange circles and the blue line segment illustrate visual elements presented on displays, while orange balls and the blue cone are AR-rendered 3D objects in the physical space.}
  \label{fig-designs}
\vspace{-5.5mm}
\end{figure}

There are four major design strategies that can support blending the representation boundary in MVs: 1) \textit{highlighting}, 2) \textit{linking}, 3) \textit{embeding}, and 4) \textit{extending}.
Figure \ref{fig-designs} illustrates detailed examples of boundary-blending designs for two views (i.e., a scatterplot on a desktop monitor and a line chart on a laptop) based on the strategies.
Moreover, Table \ref{tab-designs} gives a summary of them.

\begin{table*}[tb]
\centering
\caption{A summary of four major design strategies that can support presenting the boundary blending in MVs.}
\vspace{-1mm}
\label{tab-designs}
\resizebox{\textwidth}{!}{%
\begin{tabular}{c|c|l|l}
\textbf{Strategy} &
  \textbf{Example} &
  \multicolumn{1}{c|}{\textbf{Pros}} &
  \multicolumn{1}{c}{\textbf{Cons}} \\ \hline
\textit{Highlighting} &
  \begin{tabular}[c]{@{}c@{}}Visual \\ Salience\end{tabular} &
  \begin{tabular}[c]{@{}l@{}}- Exiting views are well-preserved.\\ - It is easy to perceive.\end{tabular} &
  \begin{tabular}[c]{@{}l@{}}- It is not efficient for views across displays with large physical gaps.\\ - It is difficult for users to check cross-boundary data relations in detail.\end{tabular} \\ \hline
\textit{Linking} &
  \begin{tabular}[c]{@{}c@{}}Visual \\ Links\end{tabular} &
  \begin{tabular}[c]{@{}l@{}}- It offers visual guidance for tracing related data across boundaries.\\ - It enables users to check cross-boundary data relations in detail.\end{tabular} &
  \begin{tabular}[c]{@{}l@{}}- Many visual links can lead to visual clutter.\\ - The readability of existing views can be interfered by visual links.\end{tabular} \\ \hline
\textit{Embedding} &
  \begin{tabular}[c]{@{}c@{}}Visual \\ Overlay\end{tabular} &
  \begin{tabular}[c]{@{}l@{}}- It allows users to maintain their focus on one view.\\ - It helps in reducing the cost of making transitions across views.\end{tabular} &
  \begin{tabular}[c]{@{}l@{}}- It impacts existing views, which may hurt their readability.\\ - It may cause visual clutter when many visual elements are embedded.\end{tabular} \\ \hline
\textit{Extending} &
  \begin{tabular}[c]{@{}c@{}}Outside-view \\ Visualization\end{tabular} &
  \begin{tabular}[c]{@{}l@{}}- It supports effectively using available space (i.e., physical space).\\ - It allows the creation of new, meaningful views outside existing views.\end{tabular} &
  \begin{tabular}[c]{@{}l@{}}- It impacts the completeness of existing views (e.g., moving elements out).\\ - It weakens the role of existing views for visual analysis.\end{tabular} \\ 
\end{tabular}%
}
\vspace{-5mm}
\end{table*}

\textit{Highlighting} (\scalerel*{\includegraphics{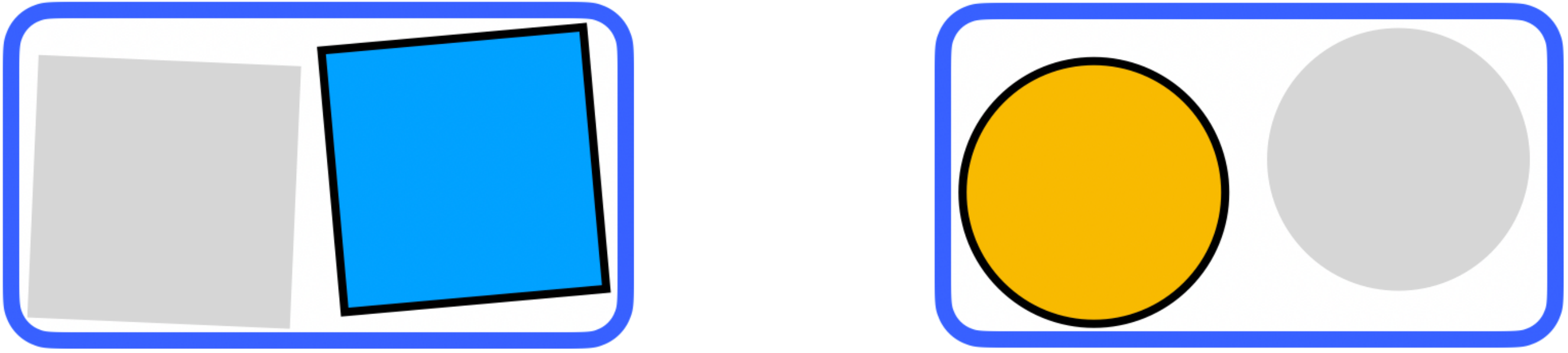}} {B}) is the most implicit design for boundary blending.
It well preserves the existing visual context of MVs, as it only controls certain visual channels (e.g., colors) to make related visual elements get salient visually.
This has been widely used in coordinated multiple views \cite{sun2021towards, munzner2014visualization, roberts2007state}.
However, as it heavily relies on the existing visual context of MVs, highlighting neither visually updates nor reforms visible boundaries (e.g., borders of views) in MVs.
Thus, the boundary blending is not explicitly shown by certain visual forms, for highlighting, which is established, instead, logically, via seeing the on and off of highlights that cross different views.   

\textit{Linking} (\scalerel*{\includegraphics{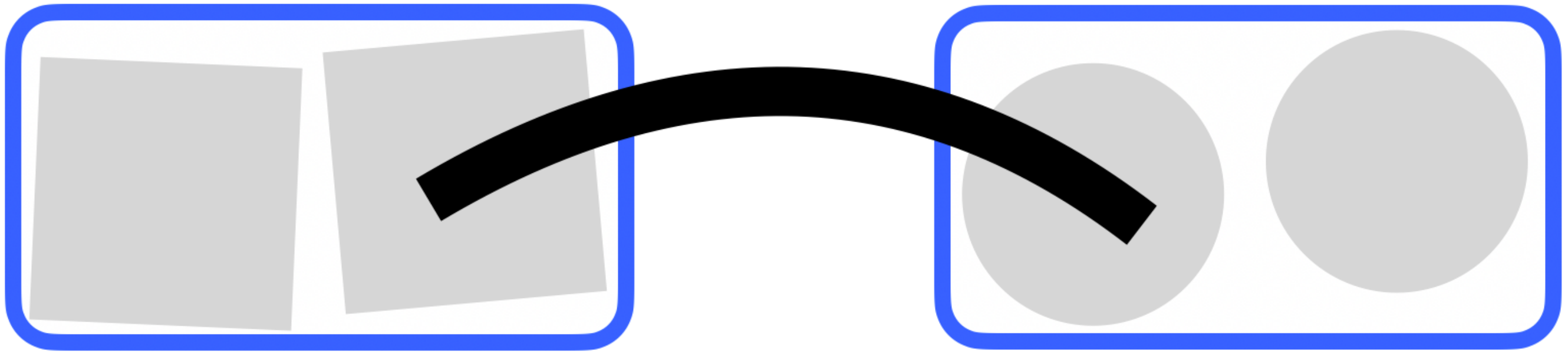}} {B}) emphasizes blending the boundary in MVs via visual links (e.g., VisLinks \cite{collins2007vislink}).
Such applied visible links serve as bridges, which enable connections among visual elements across different boundaries (i.e., from different views).
It brings a visual form that reveals the blending as linked visual elements, and users can follow such visual guidance to navigate through MVs.
Based on where these visual links locate, there are two major types of visual links: 1) \textit{on-display links}, and 2) \textit{augmented reality (AR) rendered links}.
The former relies on pixels on displays (e.g., SAViL \cite{chung2018savil}), so they look ``broken" when there are gaps between displays.
The latter uses the physical space that goes beyond displays \cite{reipschlager2020personal}.
These links offer visual affordance \cite{norman1999affordance} for the usage of boundary blending in MVs.
However, they can cause visual clutter, especially when many links are shown at the same time.

\textit{Embedding} (\scalerel*{\includegraphics{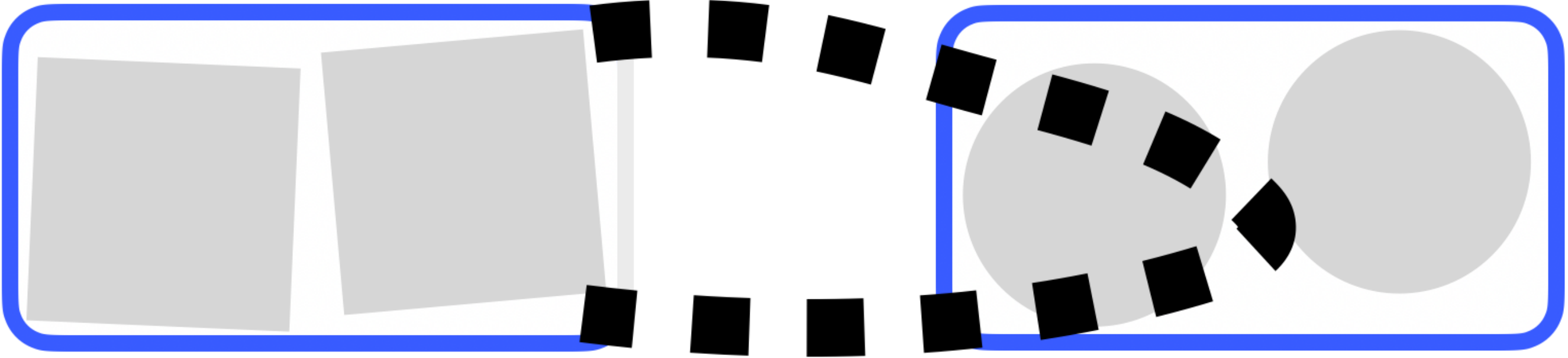}} {B}) allows shifting related visual elements from one view to another.
It pulls related data pieces (i.e., visual elements) from other views toward the data currently under investigation, so users can keep focusing on one view, without taking effort to make transitions across views (e.g., moving back and forth between two views to check related visual elements).
The shifted visual elements can be either overlaid on top of or placed neighboring their related ones.
The former can form composite visualizations \cite{javed2012exploring}. 
The latter use proximity to reveal their relations.
Such transitioning behaviors potentially pull the boundary of related views into that of the view currently under investigation.
Thus, using this strategy, the boundary blending in MVs is revealed as intertwined visual elements.
Due to the intertwinement, its generated visual representations may bring confusion to users, and also hurt the readability of existing views.

\textit{Extending} (\scalerel*{\includegraphics{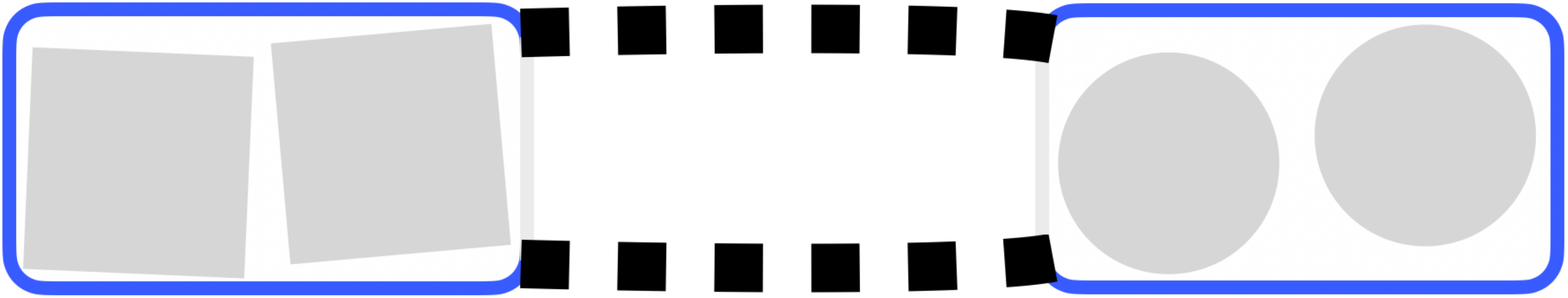}} {B}) highlights expanding the coverage of all the views in MVs.
It works similar to building tunnels, through which related visual elements from different views can be transitioned to some common area, and the coverage of views is enlarged (i.e., by using the space outside existing views).
This supports breaking and reforming the representation boundary of MVs, typically revealed as flexibly showing visual elements out of views (e.g., marks outside existing views to present cross-view data relations in SightBi \cite{sun2021sightbi} and BiSet \cite{sun2015biset}), and further creating meaningful spatializations \cite{andrews2010space}.
Similar to linking, such out of view visual elements can remain as on-display elements, or become AR rendered elements that locate in the physical space beyond displays (Figure \ref{fig-designs} (D)).
Thus, with this strategy, the boundary blending in MVs does not rely much on existing views. 
Instead, it brings new, meaningful views, corresponding to the blended parts of MVs.
Moreover, with the presence of such newly formed views, the role of existing views for visual analysis is challenged (e.g., do they just serve as a context for the new views).

In summary, from \textit{highlighting}, to \textit{linking}, \textit{embedding}, and further to \textit{extending}, MVs' representation boundaries evolve from \textit{separated}, to \textit{connected}, \textit{overlaid}, and further to \textit{blended}.
Such an evolvement more expressively reveals the boundary blending in MVs.

\section{Discussion}

We have presented three possible boundaries in MVs: data boundary, representation boundary, and semantic boundary, corresponding to three aspects of the usage of MVs: encoded data, visual representation, and interpretation (Section \ref{boundary-def}).
With them, we have discussed our notion of MVs' boundary blending and four possible designs, which support the boundary blending in MVs (highlighting, linking, embedding, and extending), particularly focusing on the representation boundary (Section \ref{sec-blending}).
While the design strategies can support visually presenting the boundary blending in MVs, possible usage of such blendings raise more open questions and challenges for the design of MVs, which may drive valuable future research directions.

\begin{figure}[tb]
  \centering
  \includegraphics[width=\columnwidth,keepaspectratio]{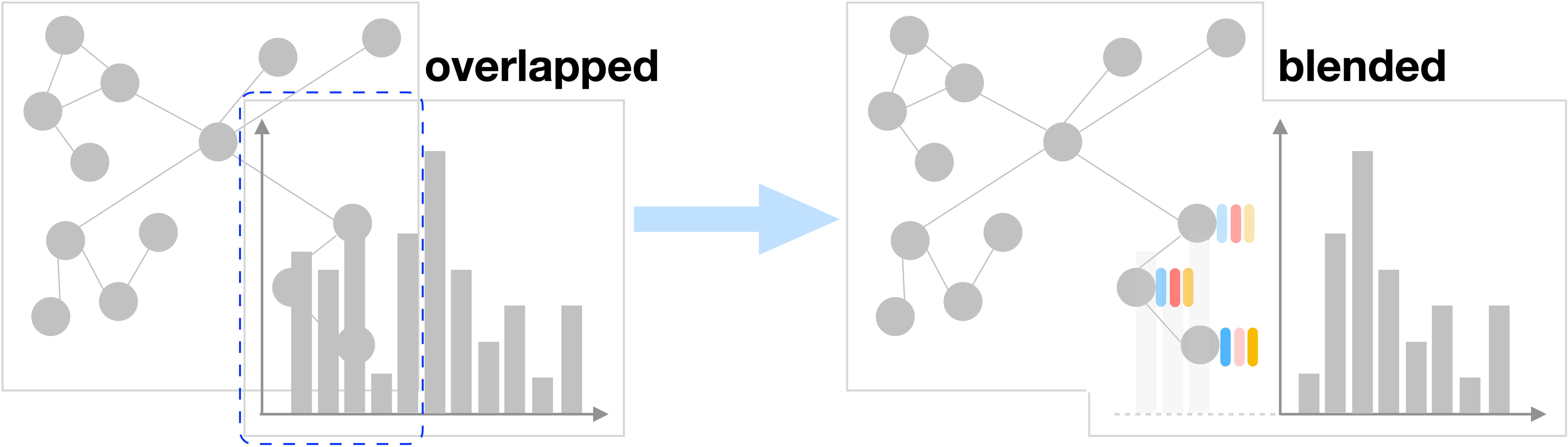}
  \vspace{-6.5mm}
  \caption{An example of changing overlapped views to a blended one. 
  The overlapped visual elements (i.e., bars) are transformed into new visual encodings (e.g., three colored mini-bars) associated with their overlapped nodes in the graph.
  }
  \label{fig-fused}
\vspace{-5.5mm}
\end{figure} 

\textit{The layout of MVs}: the boundary blending may significantly impact the design of MVs' layout.
It requires reconsidering the spatial layout of MVs from a merely perception-driven perspective \cite{zeng2023semi, chen2020composition} to a more inclusive perspective (e.g., considering both perception and the meaning of encoded data \cite{shaikh2022toward}).
For example, for perceptual needs (e.g., clear visual separation), overlap is considered harmful, which is often avoided by visualization designers and users.
However, with the concept of boundary blending, as shown in Figure \ref{fig-fused}, a harmful overlap, previously considered as visual clutters, can now be transformed into a usable, blended view.
This enriches the design space of MVs' layout by including more useful design options that may have been intentionally dropped, previously.
Future studies are needed to further develop this design space. 

\textit{The organization of MVs}: with the boundary blending, the organization of MVs needs updates accordingly.
It brings opportunities of reorganizing visual elements of MVs from a limited, local context (i.e., each individual view) to an open, global context (i.e., all views in MVs).
For example, the position of nodes in a graph is no longer constrained by the border of the graph view.
Instead, they can be flexibly put in any location (e.g., on top of a map, or in the middle of two monitors) based on certain analysis needs.
In consequence, this breaks the structure of views in the traditional sense and establishes a new way of considering the organization of MVs: shifting from a coherent chart-based structure to a meaning-oriented organization.
Such organization-oriented changes question the role of MVs in visual analysis (e.g., offering distinct visual context versus providing chains of sliced context).

\textit{The synthesis of MVs}: the boundary blending potentially serves as the basis to support future studies on the synthesis of MVs.
While effectively exploring related data across multiple views is important for information foraging, the synthesis of such data is another critical component in sensemaking \cite{pirolli2005sensemaking}.
However, how to synthesize MVs remains a key but unanswered question.
Specifically, what can we synthesize in MVs, how do we perform the synthesis accordingly, and how can we show the synthesis of MVs?
Our insights on the boundary and boundary blending in MVs may help answer them.    

\bibliographystyle{abbrv-doi}

\bibliography{ref}

\begin{thebibliography}{10}

\bibitem{car-dataset}
Data sets, interactive data visualization.
\newblock \url{https://www.idvbook.com/teaching-aid/data-sets/}.

\bibitem{acuna2019devil}
D.~Acuna, A.~Kar, and S.~Fidler.
\newblock Devil is in the edges: Learning semantic boundaries from noisy
  annotations.
\newblock In {\em Proceedings of the IEEE/CVF Conference on Computer Vision and
  Pattern Recognition}, pp. 11075--11083, 2019.

\bibitem{andrews2010space}
C.~Andrews, A.~Endert, and C.~North.
\newblock Space to think: large high-resolution displays for sensemaking.
\newblock In {\em Proceedings of the SIGCHI conference on human factors in
  computing systems}, pp. 55--64, 2010.

\bibitem{bleiholder2009data}
J.~Bleiholder and F.~Naumann.
\newblock Data fusion.
\newblock {\em ACM Computing Surveys}, 41(1):1--41, 2009.

\bibitem{castanedo2013review}
F.~Castanedo.
\newblock A review of data fusion techniques.
\newblock {\em The scientific world journal}, 2013, 2013.

\bibitem{chen2020composition}
X.~Chen, W.~Zeng, Y.~Lin, H.~M. Ai-Maneea, J.~Roberts, and R.~Chang.
\newblock Composition and configuration patterns in multiple-view
  visualizations.
\newblock {\em IEEE Transactions on Visualization and Computer Graphics},
  27(2):1514--1524, 2020.

\bibitem{chung2018savil}
H.~Chung and C.~North.
\newblock Savil: cross-display visual links for sensemaking in display
  ecologies.
\newblock {\em Personal and Ubiquitous Computing}, 22:409--431, 2018.

\bibitem{chung2015four}
H.~Chung, C.~North, S.~Joshi, and J.~Chen.
\newblock Four considerations for supporting visual analysis in display
  ecologies.
\newblock In {\em 2015 IEEE Conference on Visual Analytics Science and
  Technology}, pp. 33--40. IEEE, 2015.

\bibitem{chung2014visporter}
H.~Chung, C.~North, J.~Z. Self, S.~Chu, and F.~Quek.
\newblock Visporter: facilitating information sharing for collaborative
  sensemaking on multiple displays.
\newblock {\em Personal and Ubiquitous Computing}, 18:1169--1186, 2014.

\bibitem{collins2007vislink}
C.~Collins and S.~Carpendale.
\newblock Vislink: Revealing relationships amongst visualizations.
\newblock {\em IEEE Transactions on Visualization and Computer Graphics},
  13(6):1192--1199, 2007.

\bibitem{cornsweet2012visual}
T.~Cornsweet.
\newblock {\em Visual perception}.
\newblock Academic press, 2012.

\bibitem{coulson2001blending}
S.~Coulson and T.~Oakley.
\newblock Blending basics.
\newblock {\em Cognitive Linguistics}, 11(3-4):175--196, 2001.

\bibitem{coulson2005blending}
S.~Coulson and T.~Oakley.
\newblock Blending and coded meaning: Literal and figurative meaning in
  cognitive semantics.
\newblock {\em Journal of pragmatics}, 37(10):1510--1536, 2005.

\bibitem{dou2012leadline}
W.~Dou, X.~Wang, D.~Skau, W.~Ribarsky, and M.~X. Zhou.
\newblock Leadline: Interactive visual analysis of text data through event
  identification and exploration.
\newblock In {\em 2012 IEEE Conference on Visual Analytics Science and
  Technology}, pp. 93--102. IEEE, 2012.

\bibitem{fauconnier2003conceptual}
G.~Fauconnier and M.~Turner.
\newblock Conceptual blending, form and meaning.
\newblock {\em Recherches en communication}, 19:57--86, 2003.

\bibitem{grady1999blending}
J.~Grady, T.~Oakley, and S.~Coulson.
\newblock Blending and metaphor.
\newblock {\em Amsterdam Studies in the Theory and History of Linguistic
  Science Series 4}, pp. 101--124, 1999.

\bibitem{javed2012exploring}
W.~Javed and N.~Elmqvist.
\newblock Exploring the design space of composite visualization.
\newblock In {\em 2012 IEEE Pacific Visualization Symposium}, pp. 1--8. IEEE,
  2012.

\bibitem{johnson2010mental}
P.~N. Johnson-Laird.
\newblock Mental models and human reasoning.
\newblock {\em Proceedings of the National Academy of Sciences},
  107(43):18243--18250, 2010.

\bibitem{ko2012marketanalyzer}
S.~Ko, R.~Maciejewski, Y.~Jang, and D.~S. Ebert.
\newblock Marketanalyzer: An interactive visual analytics system for analyzing
  competitive advantage using point of sale data.
\newblock In {\em Computer Graphics Forum}, vol.~31, pp. 1245--1254. Wiley
  Online Library, 2012.

\bibitem{li2021multitask}
A.~Li, L.~Jiao, H.~Zhu, L.~Li, and F.~Liu.
\newblock Multitask semantic boundary awareness network for remote sensing
  image segmentation.
\newblock {\em IEEE Transactions on Geoscience and Remote Sensing}, 60:1--14,
  2021.

\bibitem{marin2019efficient}
D.~Marin, Z.~He, P.~Vajda, P.~Chatterjee, S.~Tsai, F.~Yang, and Y.~Boykov.
\newblock Efficient segmentation: Learning downsampling near semantic
  boundaries.
\newblock In {\em Proceedings of the IEEE/CVF International Conference on
  Computer Vision}, pp. 2131--2141, 2019.

\bibitem{meng2020survey}
T.~Meng, X.~Jing, Z.~Yan, and W.~Pedrycz.
\newblock A survey on machine learning for data fusion.
\newblock {\em Information Fusion}, 57:115--129, 2020.

\bibitem{munzner2014visualization}
T.~Munzner.
\newblock {\em Visualization analysis and design}.
\newblock CRC press, 2014.

\bibitem{norman1999affordance}
D.~A. Norman.
\newblock Affordance, conventions, and design.
\newblock {\em interactions}, 6(3):38--43, 1999.

\bibitem{north2006toward}
C.~North.
\newblock Toward measuring visualization insight.
\newblock {\em IEEE computer graphics and applications}, 26(3):6--9, 2006.

\bibitem{pirolli2005sensemaking}
P.~Pirolli and S.~Card.
\newblock The sensemaking process and leverage points for analyst technology as
  identified through cognitive task analysis.
\newblock In {\em Proceedings of International Conference on Intelligence
  Analysis}, vol.~5, pp. 2--4. McLean, VA, USA, 2005.

\bibitem{reipschlager2020personal}
P.~Reipschlager, T.~Flemisch, and R.~Dachselt.
\newblock Personal augmented reality for information visualization on large
  interactive displays.
\newblock {\em IEEE Transactions on Visualization and Computer Graphics},
  27(2):1182--1192, 2020.

\bibitem{roberts2007state}
J.~C. Roberts.
\newblock State of the art: Coordinated \& multiple views in exploratory
  visualization.
\newblock In {\em International Conference on Coordinated \& Multiple Views in
  Exploratory Visualization}, pp. 61--71. IEEE, 2007.

\bibitem{shaikh2022toward}
A.~R. Shaikh, D.~Koop, H.~Alhoori, and M.~Sun.
\newblock Toward systematic design considerations of organizing multiple views.
\newblock In {\em 2022 IEEE Visualization and Visual Analytics (VIS)}, pp.
  105--109. IEEE, 2022.

\bibitem{jigsaw}
J.~Stasko, C.~G{\"o}rg, and Z.~Liu.
\newblock Jigsaw: supporting investigative analysis through interactive
  visualization.
\newblock {\em Information Visualization}, 7(2):118--132, 2008.

\bibitem{steinberger2011context}
M.~Steinberger, M.~Waldner, M.~Streit, A.~Lex, and D.~Schmalstieg.
\newblock Context-preserving visual links.
\newblock {\em IEEE Transactions on Visualization and Computer Graphics},
  17(12):2249--2258, 2011.

\bibitem{sultanum2022chartwalk}
N.~Sultanum, F.~Naeem, M.~Brudno, and F.~Chevalier.
\newblock Chartwalk: Navigating large collections of text notes in electronic
  health records for clinical chart review.
\newblock {\em IEEE Transactions on Visualization and Computer Graphics},
  29(1):1244--1254, 2022.

\bibitem{sun2015biset}
M.~Sun, P.~Mi, C.~North, and N.~Ramakrishnan.
\newblock Biset: Semantic edge bundling with biclusters for sensemaking.
\newblock {\em IEEE transactions on visualization and computer graphics},
  22(1):310--319, 2015.

\bibitem{sun2021towards}
M.~Sun, A.~Namburi, D.~Koop, J.~Zhao, T.~Li, and H.~Chung.
\newblock Towards systematic design considerations for visualizing cross-view
  data relationships.
\newblock {\em IEEE Transactions on Visualization and Computer Graphics},
  28(12):4741--4756, 2021.

\bibitem{sun2021sightbi}
M.~Sun, A.~R. Shaikh, H.~Alhoori, and J.~Zhao.
\newblock Sightbi: Exploring cross-view data relationships with biclusters.
\newblock {\em IEEE Transactions on Visualization and Computer Graphics},
  28(1):54--64, 2021.

\bibitem{wilkinson2012grammar}
L.~Wilkinson.
\newblock {\em The grammar of graphics}.
\newblock Springer, 2012.

\bibitem{bitextract}
X.~Yue, X.~Shu, X.~Zhu, X.~Du, Z.~Yu, D.~Papadopoulos, and S.~Liu.
\newblock Bitextract: Interactive visualization for extracting bitcoin exchange
  intelligence.
\newblock {\em IEEE Transactions on Visualization and Computer Graphics},
  25(1):162--171, 2019. doi: {{%
10\hspace{.1pt}\discretionary{.}{%
}{.}\hspace{.4pt}1109\discretionary{/}{%
}{/}TVCG\hspace{.1pt}\discretionary{.}{%
}{.}\hspace{.4pt}2018\hspace{.1pt}\discretionary{.}{%
}{.}\hspace{.4pt}2864814}}


\bibitem{zeng2023semi}
W.~Zeng, X.~Chen, Y.~Hou, L.~Shao, Z.~Chu, and R.~Chang.
\newblock Semi-automatic layout adaptation for responsive multiple-view
  visualization design.
\newblock {\em IEEE Transactions on Visualization and Computer Graphics}, 2023.

\bibitem{zhang2015visualizing}
H.~Zhang, M.~Sun, D.~Yao, and C.~North.
\newblock Visualizing traffic causality for analyzing network anomalies.
\newblock In {\em Proceedings of the ACM International Workshop on Security and
  Privacy Analytics}, pp. 37--42, 2015.

\end{thebibliography}
\end{document}